# Influence of the Gravitational Darkening Effect on the Spectrum of a Hot, Rapidly Rotating Neutron Star. II. Iron Lines

Agnieszka Majczyna[1], Jerzy Madej[2], Agata Różańska[3], and Mirosław Należyty[2]
[1] National Centre for Nuclear Research, ul. Andrzeja Sołtana 7, 05-400 Otwock, Poland; agnieszka.majczyna@ncbj.gov.pl
[2] Astronomical Observatory, University of Warsaw, Al. Ujazdowskie 4, 00-478 Warszawa, Poland
[3] Nicolaus Copernicus Astronomical Centre, Polish Academy of Sciences, ul. Bartycka 18, 00-716 Warszawa, Poland

## Abstract

Rapidly rotating neutron stars are similar to highly flattened ellipsoids. Observed spectra of flattened stars must exhibit effects of nonspherical shape and gravitational darkening. We examined in detail the influence of both effects on the observed central energies and profiles of lines of highly ionized iron, Fe XXV and Fe XXVI. We note that the gravitational darkening effect does not change the central energy of lines. Most importantly, spectra of neutron stars that rotate with different frequencies and are seen at various inclination angles differ significantly. The appearance and the depth of lines strongly depend on the parameters, like the inclination angle of the star or the frequency of the star rotation. In this paper we clearly show that the gravitational darkening effect should be included in realistic models of the atmospheres of the neutron stars.

*Unified Astronomy Thesaurus concepts:* Neutron stars (1108); Radiative transfer equation (1336)

## 1. Introduction

Spectral lines from neutron star surfaces are one of very promising features for direct probing of the properties of these objects like mass and radius. These two parameters allow us to determine or constrain the equation of state and look into the neutron star interior. Spectral lines could be used for the gravitational redshift measurements and therefore for compactness ratio determination. A separate issue is the line identification and place where these lines are formed.

Cottam et al. (2002) reported detection of narrow absorption lines in the burst spectra of the low-mass X-ray binary EXO 0748-676. These lines were interpreted as Fe XXVI, Fe XXV $n = 2$–3, and O VII $n = 1$–2 transitions with the redshift $z = 0.35$. Up to now, this discovery was not confirmed, but the redshift $z = 0.35$ is widely used as the canonical value. If the spin frequency of the neutron star is known, then the width of a spectral line allows for measurements of the radius of the neutron star (Özel & Psaltis 2003). For this reason, modeling of spectral lines that include all possible effects are very important for the determination of the mass and the radius of the neutron star and therefore for constraining the equation of state of the superdense matter that builds up the neutron star.

The presence of iron in the atmosphere of a neutron star is still under debate. There are papers that report about iron lines in the spectrum of low-mass X-ray binaries, but those features are interpreted as lines from the accretion disk or hot boundary layers between a neutron star atmosphere and the disk (see, e.g., Bhattacharyya & Strohmayer 2007). However, there are papers that suggested the presence of iron lines in the atmosphere of a neutron star (see, e.g., Li et al. 2018).

The presence of spectral features in the spectra of neutron stars and their properties strongly depends on the chemical composition of the atmosphere, the effective temperature, the surface gravity, and other parameters. The chemical composition of the atmosphere is poorly known, but there are many issues that indicate what species are present in atmospheres of neutron stars in X-ray bursters. For example, recently Sharma et al. (2022) presented the broadband spectrum of low-mass X-ray binary 2S 0921-63 detected by Suzaku. In this paper authors reported observation of the broad emission line around 6.7 keV. This feature can be identified with the unreddened resonance line of the highly ionized iron Fe XXV. It is not excluded that the line is related to the atmosphere of a neutron star.

In our previous paper, we computed theoretical spectra of rotationally flattened neutron stars with hydrogen–helium atmosphere (see Majczyna et al. 2022). We investigated the influence of the gravitational darkening effect on the spectra of an oblate neutron star. In our theoretical models, we assumed various values of the effective temperature $T_{\text{eff}}$, the surface gravity $\log(g)$, rotational velocity (dimensionless angular velocity $\bar{\Omega}^2$), and inclination angle $i$ (angle between the equator and direction to the observer). We also compare these spectra with the spectra of the spherical star. All spectra of flattened stars significantly differ from spectra of the spherical star, and their shape strongly depends on the dimensionless angular velocity and inclination angle for fixed temperature and surface gravity. We clearly showed that gravitational darkening effects should be included in realistic models of the atmospheres of the neutron stars.

## 2. Our Model

### 2.1. Distribution of the Surface Gravity and the Effective Temperature

In our investigations, we assume that neutron stars are flattened by a rapid rotation and their atmospheres are very hot. Such neutron stars reside in low-mass X-ray binaries that show type I X-ray bursts. Before calculation of stellar atmosphere models, we should define the geometry of the neutron star and next the distribution of the surface temperature and surface gravity. These two latter values are input parameters for model atmospheres calculated by the ATM24 code. We used a spherical coordinate system where $\theta = 0°$ denotes the latitude of the equator.

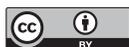

We assumed that a flattened neutron star could be described as ellipsoid with the equatorial radius $R_{eq}$ defined as

$$R(\theta) = R_{eq}\sqrt{1 - e^2\cos(90° - \theta)}, \quad (1)$$

where $\theta$ is the latitude in spherical coordinates, $R_{eq}$ is the equatorial radius, and $e$ is the eccentricity.

The eccentricity depends on the equation of state and on the dimensionless angular velocity of star rotation ($\bar{\Omega}^2$). Formulae connected these two parameters were proposed by various authors (see, e.g., AlGendy & Morsink 2014 and Silva et al. 2021). In our analysis we use the approximate relation from Silva et al. (2021), suitable for fast rotation. It should be emphasized that we assumed two values of the dimensionless angular velocities ($\bar{\Omega}^2 = 0.30$ and $0.60$) corresponding to the fast-rotation regime. Here we quoted Equation (19) from Silva et al. (2021) using variables defined in our work:

$$e = 0.251 + 0.935\bar{\Omega}^2 + 0.709x + 0.030\bar{\Omega}^2 x \\ - 0.472\bar{\Omega}^4 - 2.427x^2, \quad (2)$$

where $x = M/R_{eq}$ and $\bar{\Omega}$ is dimensionless angular velocity defined as

$$\bar{\Omega} = \Omega\left(\frac{R_{eq}^3}{GM}\right)^{1/2}. \quad (3)$$

In the case of a flattened neutron star, the effective surface gravity is a local value, which means that gravity depends on the latitude angle $g(\theta)$. The highest gravity is on poles and decrease to the lowest value on the equator. AlGendy & Morsink (2014) presented formulae on the local surface gravity in two regimes— fast and slow rotation of the neutron star. We use Equation (50) from this paper, but we rewrite this equation to the coordinate system where $\theta$ is the latitude angle. AlGendy & Morsink (2014) used the colatitude angle ($\theta' = 90° - \theta$):

$$g(\theta)/g_0 = 1 + (c_e\bar{\Omega}^2 + d_e\bar{\Omega}^4 + f_e\bar{\Omega}^6)\sin^2(90° - \theta) \\ + (c_p\bar{\Omega}^2 + d_p\bar{\Omega}^4 + f_p\bar{\Omega}^6 - d_{60}\bar{\Omega}^4)\cos^2 \\ \times (90° - \theta) + d_{60}\bar{\Omega}^4\cos(90° - \theta), \quad (4)$$

where $g_0$ is the gravity on the surface of the nonrotating star and is equal to

$$g_0 = \frac{GM}{R_{eq}^2}\left(1 - \frac{2GM}{R_{eq}c^2}\right)^{-1/2}. \quad (5)$$

Coefficients $c_e$, $d_e$, etc. were defined in AlGendy & Morsink (2014) as follows:

$$\begin{aligned} c_e &= -0.791 + 0.776x, \\ d_e &= -1.315x + 2.431x^2, \\ f_e &= -1.172x, \\ c_p &= 1.138 - 1.431x, \\ d_p &= 0.653x - 2.864x^2, \\ f_p &= 0.975x, \\ d_{60} &= 13.67x - 27.13x^2. \end{aligned} \quad (6)$$

According to the von Zeipel theorem, the effective temperature is a function of the local surface gravity $T(g(\theta))$, and it can be expressed as

$$T(\theta) = T_{eff}\left(\frac{g(\theta)}{g_0}\right)^{1/4}, \quad (7)$$

where $T(\theta)$ is the local temperature, $T_{eff}$ is the effective temperature of the spherical star, $g(\theta)$ is the local surface gravity, and $g_0$ is the gravity of the spherical star.

In our investigations, we assume that the gravitational darkening exponent (GDE) is equal to 1/4 because we approximate the hot atmosphere of the neutron star by the radiative atmosphere. When the atmosphere is not radiative, the GDE has a different value for nonradiative or not fully radiative envelope as was shown by Lucy (1967). The author also showed that for convective envelopes, this exponent has a value ~0.08. Claret (2021) presented self-consistent calculations of the GDE for DA and DB white dwarfs and showed that GDE depends not only on the temperature but also on the opacity. We do not calculate GDE, but we assume its value.

Figure 1 shows the dependency of the effective temperature (left panel) and the effective surface gravity (right panel) on the latitude angle $\theta$ for two values of dimensionless angular velocities $\bar{\Omega}^2 = 0.30$ (blue line) and 0.60 (red line). For both values of $\bar{\Omega}^2$, the highest values of the effective temperature and the surface gravity are on the pole of the distorted star, and in contrast, the lowest values are at the equator ($\theta = 0°$). Importantly, the surface gravity on the pole of a distorted star is not equal to the surface gravity of the spherical star.

The influence of various parameters, e.g., GDE on the effective temperature and the gravity distribution over the surface of flattened star was presented in the previous paper (Majczyna et al. 2022); therefore, we omit this issue in the present paper.

### 2.2. Model Atmospheres

We calculated the local model of atmospheres and intensity spectra at various points on the surface of the oblate star. Models were calculated by our ATM24 code after prior determination of the local effective temperature and the gravity using equations of Section 2.1.

The ATM24 program iteratively calculates the distribution of temperature and gas pressure, density, and other variables in a plane-parallel atmosphere in hydrostatic and radiative equilibrium. The code assumes relativistic distribution of the thermal velocities of atoms and free electrons. The code assumes that gas in the star's atmosphere is in the state of local thermodynamic equilibrium. Influence of the magnetic field is neglected here.

Equations and calculation methods of the ATM24 code have been described in detail in earlier papers (Madej 1991; Madej et al. 2004; Majczyna et al. 2005), and the accuracy of our model was presented in Madej et al. (2017) and Vincent et al. (2018).

The equation of transfer was adopted from Pomraning (1973; see also Sampson 1959) and has the following

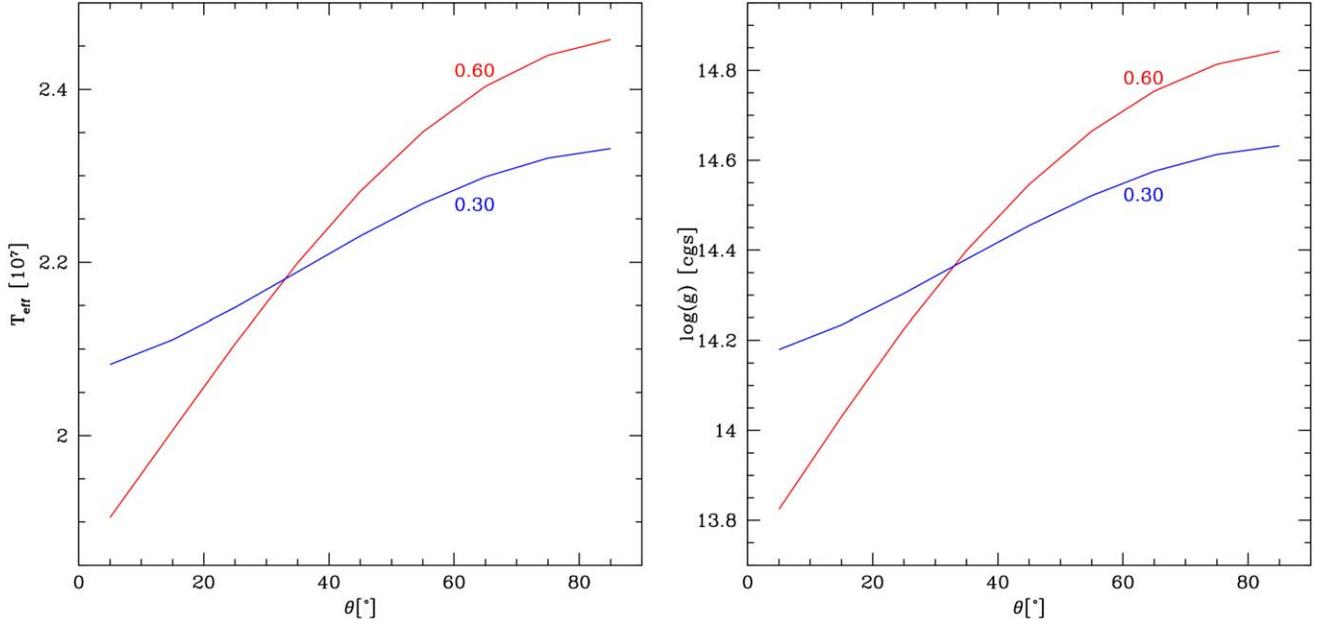

**Figure 1.** Dependency of the effective temperature (left panel) and the effective surface gravity (right panel) on the latitude angle $\theta$ for various values of dimensionless angular velocities. We assumed the effective temperature of spherical star $T_{\rm eff} = 2.20 \times 10^7$ K, logarithm of the surface gravity of a spherical star $\log(g) = 14.40$, GDE $\beta = 0.25$, and two values of dimensionless angular velocities $\bar{\Omega}^2 = 0.30$ (blue line) and 0.60 (red line).

quadratic form:

$$\mu \frac{dI_\nu}{d\tau_\nu} = I_\nu - \frac{k_\nu}{k_\nu + \sigma_\nu} B_\nu - \left(1 - \frac{k_\nu}{k_\nu + \sigma_\nu}\right) J_\nu$$
$$+ \left(1 - \frac{k_\nu}{k_\nu + \sigma_\nu}\right) J_\nu \int_0^\infty \Phi(\nu, \nu') \left(1 + \frac{c^2}{2h\nu'^3} J_{\nu'}\right) d\nu' +$$
$$- \frac{k_\nu}{k_\nu + \sigma_\nu} \left(1 + \frac{c^2}{2h\nu^3} J_\nu\right)$$
$$\times \int_0^\infty \Phi(\nu, \nu') J_{\nu'} \left(\frac{\nu}{\nu'}\right)^3 \exp\left[-\frac{h(\nu - \nu')}{kT}\right] d\nu', \qquad (8)$$

where $k_\nu$ and $\sigma_\nu$ denote coefficients of absorption and electron scattering, respectively. $I_\nu$ is the energy-dependent specific intensity, $J_\nu$ is the mean intensity of radiation, and $z$ is the geometrical depth in the atmosphere. The ATM24 code calculates a rich set of continuum bound-free and free-opacities $k_\nu$ of hydrogen, helium, and iron in all stages of ionization and line opacities of selected (up to 18) lines of Fe XXV and Fe XXVI ions.

We applied the exact angle-averaged redistribution function $\Phi(\nu, \nu')$ from Nagirner & Poutanen (1993, 1994), and the Compton scattering cross section $\sigma(\nu \rightarrow \nu', \mathbf{n} \cdot \mathbf{n}')$ was computed according to Madej et al. (2017).

In hot atmospheres of neutron stars (X-ray bursters), Compton scattering plays a crucial role in the formation of the emergent spectrum. Therefore, our model deals with this process very carefully. Influence of Compton scattering on the spectrum was described in our previous papers (e.g., Majczyna et al. 2005). In Majczyna et al. (2005), we calculated spectra of hot neutron stars having atmospheres containing iron of two different abundances (solar abundance and 100 times solar). Since model spectra presented in Majczyna et al. (2005) were emitted from the unit surface of a star, they can be simply rescaled only to spectra of spherical stars. Spectra of rotating, flattened stars can be only obtained by numerical integration of the local intensity spectra over visible part of the stars' surface taking into account both the law of limb darkening/brightening and the effect of gravitational darkening.

**Table 1**
Fe XXVI Doublet Lines Were Replaced Here by Singlet Lines at the Averaged Wavelength

| Ion | Transition | Multiplet | $\lambda$ (Å) | keV | $g_i$ | $f_{ik}$ |
|---|---|---|---|---|---|---|
| Fe XXV | $1s^2$–$1s2p$ | $^1S$ –$^1P°$ | 1.8504 | 6.70005 | 1 | 0.798 |
| Fe XXV | $1s^2$–$1s3p$ | $^1S$ –$^1P°$ | 1.5732 | 7.88103 | 1 | 0.156 |
| Fe XXV | $1s^2$–$1s4p$ | $^1S$ –$^1P°$ | 1.4946 | 8.29549 | 1 | 0.0579 |
| Fe XXV | $1s^2$–$1s5p$ | $^1S$ –$^1P°$ | 1.4608 | 8.48743 | 1 | 0.0278 |
| Fe XXVI | $1s$ – $2p$ | $^2S$ –$^2P°$ | 1.7798 | 6.96620 | 2 | 0.416 |
| Fe XXVI | $1s$ – $3p$ | $^2S$ –$^2P°$ | 1.5028 | 8.25021 | 2 | 0.0790 |
| Fe XXVI | $1s$ – $4p$ | $^2S$ –$^2P°$ | 1.4251 | 8.52068 | 2 | 0.0290 |
| Fe XXVI | $1s$ – $5p$ | $^2S$ –$^2P°$ | 1.3918 | 8.97267 | 2 | 0.0139 |

### 2.3. Iron Line Opacities

In this work we appended opacity of the four lowest lines of the fundamental series of helium-like and hydrogen-like iron ions, respectively (Fe XXV and Fe XXVI). Those iron ions are most abundant in hot atmospheres of neutron stars of the effective temperatures about $2.0 \times 10^8$ K.

Central energies, $g_i$, and $f$ values for iron lines of both ions were obtained from tables of the Opacity Project (see Seaton et al. 1992) and from Verner et al. (1996) and are presented in Table 1. Central energies of those iron lines fall in the energy range from 6.7 keV up to 9.2 keV.

The line absorption coefficient (for one iron atom) was computed according to the standard formula:

$$\kappa(\Delta\nu) = \frac{\pi e^2}{m_e c} f_{ik} \, \psi(\Delta\nu), \qquad (9)$$

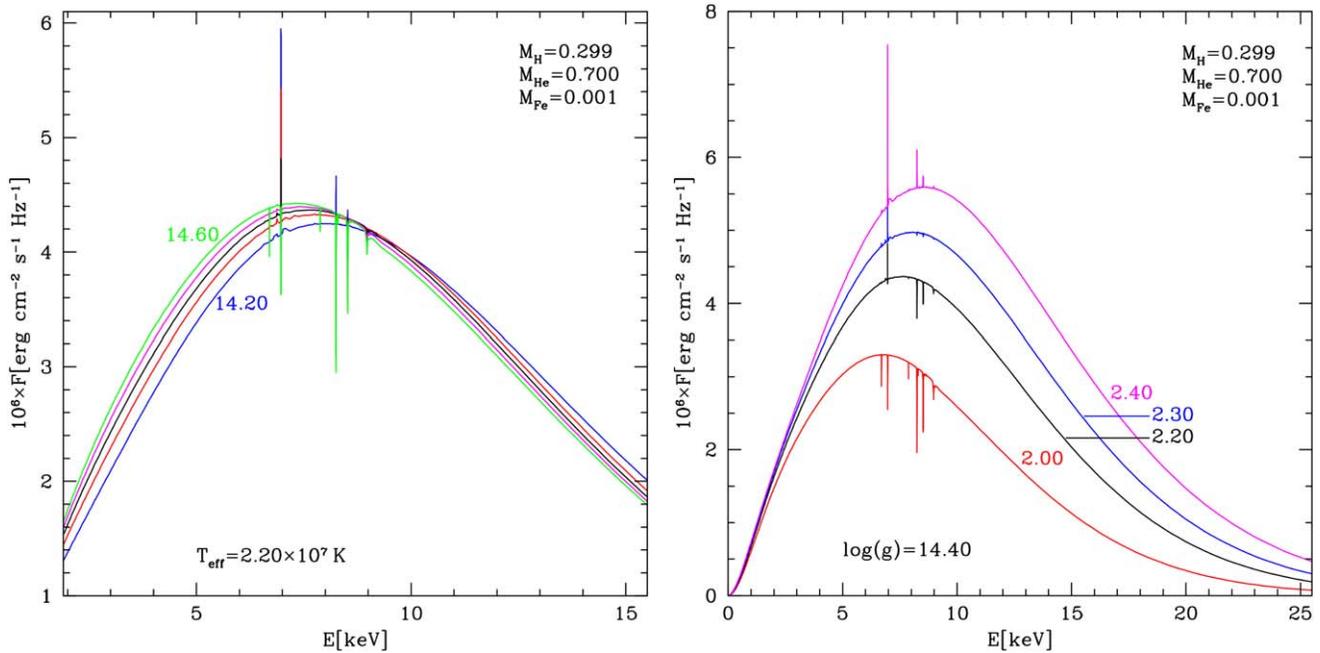

**Figure 2.** Theoretical spectra of the unit surface calculated by the ATM24 code for different values of the effective temperatures and surface gravities. In the left panel, we show spectra for the fixed effective temperature $T_{\rm eff} = 2.20 \times 10^7$ K and various surface gravities from $\log(g) = 14.20$ (cgs) to $\log(g) = 14.60$ (cgs). In the right panel, we show spectra for the fixed surface gravity $\log(g) = 14.40$ (cgs) and various effective temperature from $T_{\rm eff} = 2.00 \times 10^7$ K to $T_{\rm eff} = 2.60 \times 10^7$ K.

where $\Delta\nu$ denotes frequency difference relative to the line center (in Hz) and $\psi$ is the line-broadening profile, normalized to unity

$$\int_{-\infty}^{\infty} \psi(\Delta\nu) d\nu = 1. \tag{10}$$

Our code carefully computed broadening of each line as the convolution of natural, thermal (Doppler), and pressure (Stark) broadening (see, e.g., Mihalas 1978). In our approach, we neglected additional broadening mechanisms like magnetic broadening or Doppler broadening caused by turbulence, rotation, or bulk motion of matter. Partial profiles of Fe lines jointly broadened by thermal and natural mechanisms were approximated by the (properly rescaled) Voigt function, $\pi^{-1/2} H(a, v)/\Delta\nu_D$. Both the Doppler width $\Delta\nu_D$ for iron ions and the damping parameters $a_{ik}$ for individual lines were estimated by ATM24 code at each level using values of the local temperature $T$ and the local energy-dependent Planck function, $B_\nu(T)$. We adopted the following approximations to account for this type of line broadening:

1. Fe XXV lines: quadratic Stark approximation by Cowley (1971) and Rauch et al. (2008);
2. Fe XXVI 1s–2p resonance line: linear Stark approximate formulae by Griem (1974), who considered both quasi-static ion and impact-free electron contributions; and
3. Fe XXVI higher lines: linear Stark approximation by van Dien (1949; see also Rauch et al. 2008).

Final line-broadening profiles $\psi$ were obtained by numerical convolution of the pressure (Stark) broadening function with the Voigt function.

### 2.4. Computational Results

In this work, we assume that the reference spherical neutron star has the effective temperature $T_{\rm eff} = 2.20 \times 10^7$ K and the surface gravity $\log(g) = 14.40$ (cgs). Chemical composition of the atmosphere was assumed as a mixture of hydrogen $M_{\rm H} = 0.299$, helium $M_{\rm He} = 0.700$, and iron $M_{\rm Fe} = 0.001$. Chemical composition of the neutron star atmosphere is poorly known. There is no doubt that these atmospheres contain hydrogen and helium in different, exactly unknown proportions (e.g., Galloway et al. 2006). We are aware that the chemical composition assumed in this paper could not correspond to the real chemical composition of the neutron star atmospheres in X-ray bursters. But our conclusions are valid for iron lines as well as for lines from other elements that could be present in the atmosphere. We want to emphasize that our conclusions can be extended to normal or evolved distorted stars in close binaries. Iron lines are the only example that shows how the gravitational darkening effect influences spectral lines.

In Figure 2, we show our theoretical spectra calculated by the ATM24 code for the assumed chemical composition and a few values of the surface gravity (left panel) and a few values of the effective temperature (right panel). These spectra are emitted by the unit surface on the star. In all these spectra, iron lines are present, but their intensity is different. For the highest effective temperature (see right panel) and the lowest surface gravity (left panel), iron lines are very weak, and in contrast, for the lowest effective temperatures and the highest gravities, lines are very prominent and all of them appear in absorption.

### 3. Results

We calculated a small grid of theoretical spectra of rotating neutron stars (32 models). We divided the stars into $18 \times 36$ points in latitude and azimuthal angles. In our model, the atmosphere is radiative; therefore, GDE $\beta = 1/4$ was assumed. We assumed also parameters of the rotating star—the effective temperature of the spherical star $T_{\rm eff} = 2.20 \times 10^7$ K, surface gravity of the nonrotating star $\log(g) = 14.40$ (cgs), chemical composition ($M_{\rm H} = 0.299$, $M_{\rm He} = 0.700$, and $M_{\rm Fe} = 0.001$), and mass-to-radius ratio $x = 0.195$. Our theoretical spectra were calculated for different values of the dimensionless angular velocities $\bar{\Omega}^2 = 0.30$ and $0.60$ and inclination angles from $i = 0°$ up to $90°$ with the step $\Delta i = 10°$.

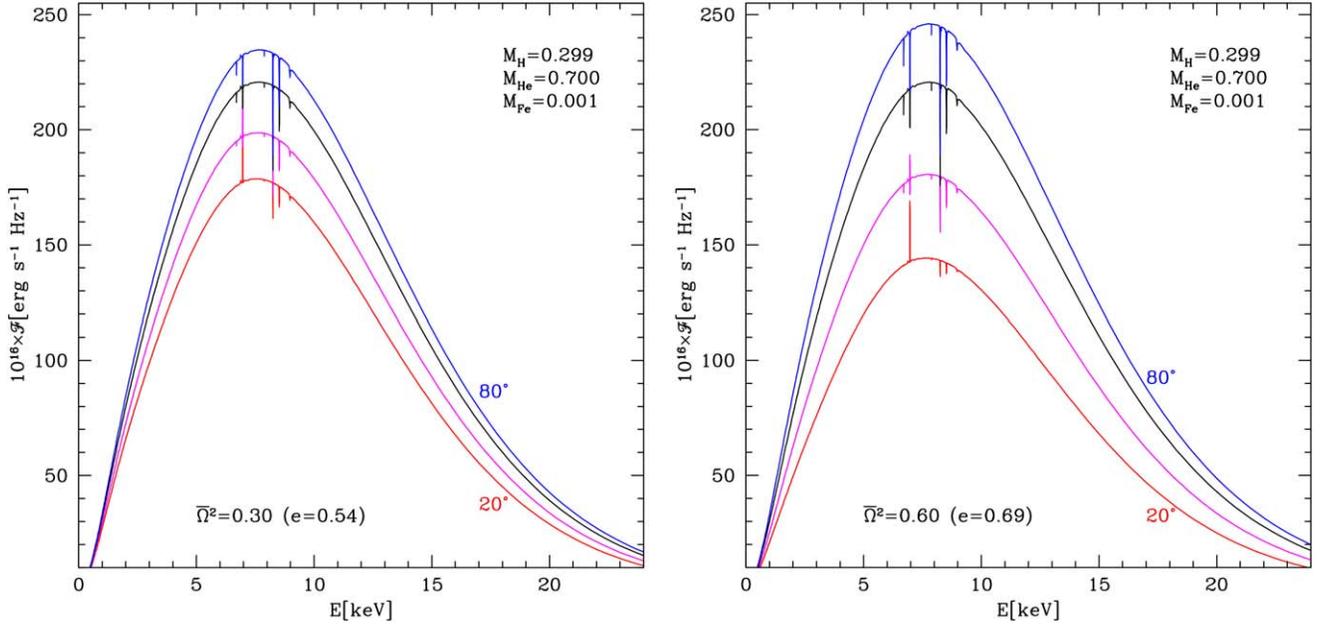

**Figure 3.** Theoretical spectra of the distorted neutron star integrated over the whole surface for various values of the inclination angle. In the left panel, we show spectra for fixed value of the effective temperature of the spherical star $T_{\text{eff}} = 2.20 \times 10^7$ K, surface gravity of the spherical star $\log(g) = 14.40$ (cgs), dimensionless angular velocity $\bar{\Omega}^2 = 0.30$, and various values of the inclination angle from $i = 20°$ to $i = 80°$. In the right panel, we show spectra for the same as previous values of $T_{\text{eff}}$, $\log(g)$, dimensionless angular velocity $\bar{\Omega}^2 = 0.60$, and various values of the inclination angle from $i = 20°$ to $i = 80°$.

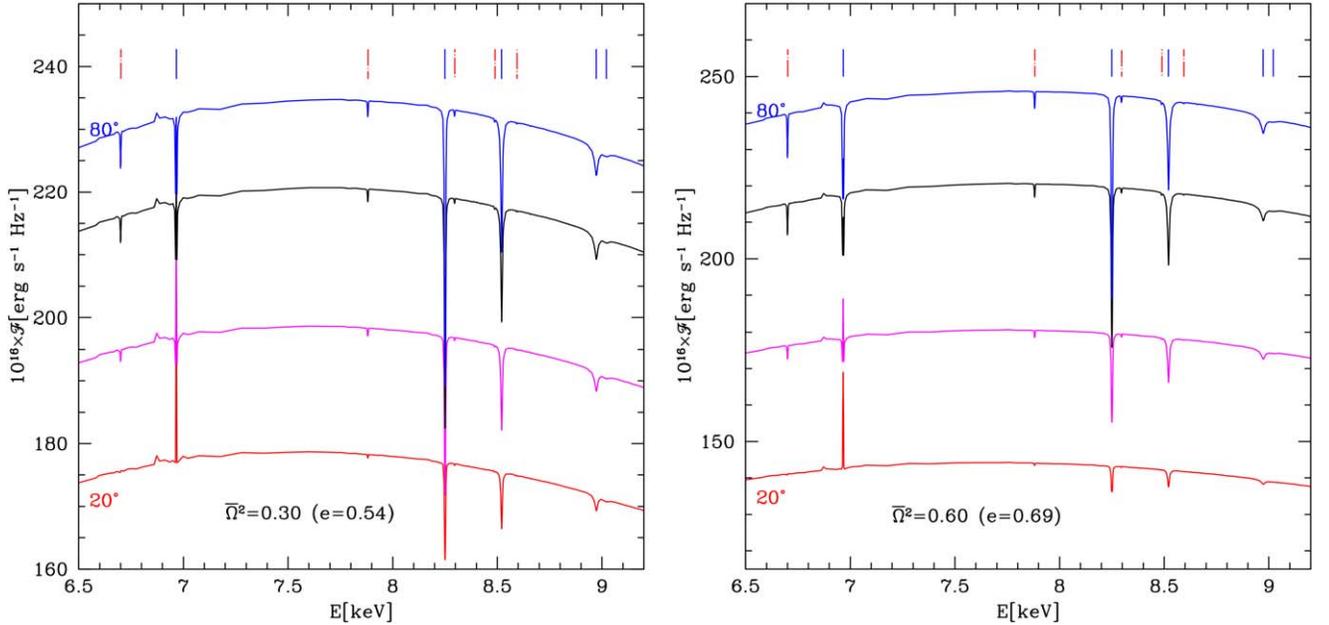

**Figure 4.** Theoretical spectra of the rotating neutron star are the same as in Figure 3. The area where spectral lines are present was magnified. By red dotted–dashed lines we mark the central energies of the fundamental series of helium-like iron, whereas blue solid lines mark spectral lines from the fundamental series of hydrogen-like iron.

Figure 3 shows theoretical spectra of the rotating neutron star integrated over the whole surface as could be seen by distant observer. We assumed the effective temperature of the spherical star $T_{\text{eff}} = 2.20 \times 10^7$ K, surface gravity of the spherical star $\log(g) = 14.40$ (cgs), GDE $\beta = 1/4$, two values of the dimensionless angular velocities $\bar{\Omega}^2 = 0.30$ (left panel) and $\bar{\Omega}^2 = 0.60$, and four values of the inclination angles. In our calculations, the eccentricity of the star is connected with $\bar{\Omega}^2$ by Equation (2) according to the formula given by Silva et al. (2021). Therefore, in Figure 3, in brackets we show the eccentricity $e$ calculated from the mentioned relation. For both assumed speeds of the rotation, iron lines are clearly visible, but for lower $i$, the lines are weaker than for higher values of the inclination angles. An enlarged area of Figure 4, where spectral lines are present, is shown in Figure 5. By dotted–dashed red lines we denote central energies of fundamental series spectral lines from helium-like iron and by solid lines we denote the central energies from hydrogen-like iron.

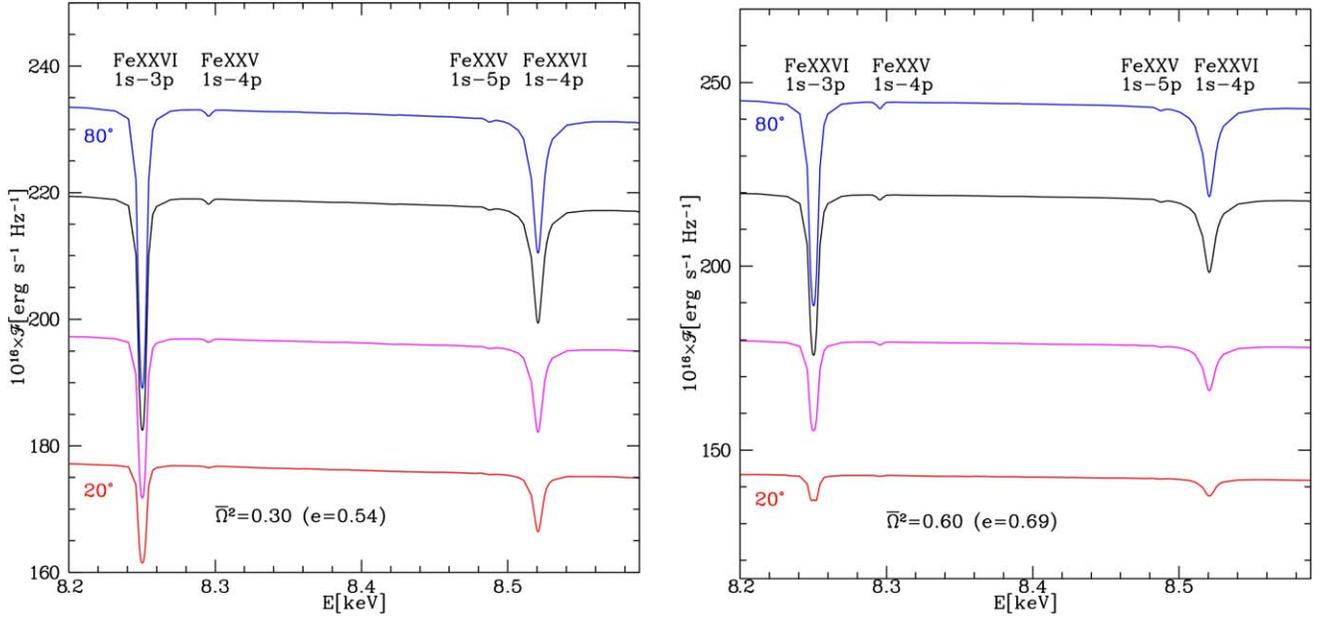

**Figure 5.** Theoretical spectra of the rotating neutron star around most prominent iron line.

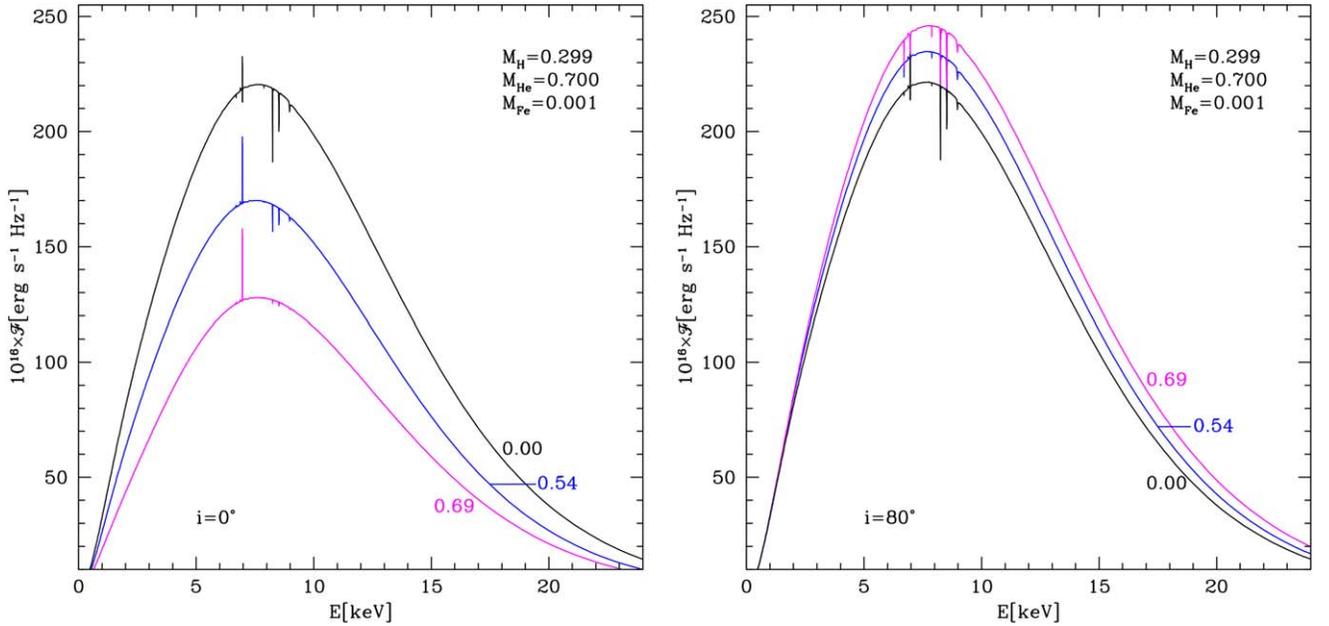

**Figure 6.** Theoretical spectra of the neutron star integrated over the whole surface for various eccentricities. We assumed the effective temperature of the spherical star $T_{\mathrm{eff}} = 2.20 \times 10^7$ K, surface gravity of the spherical star $\log(g) = 14.40$ (cgs), and two inclination angles $i = 0°$ (left panel) and $i = 80°$ (right panel) and three various eccentricities of the neutron star $e = 0.0$ ($\bar{\Omega}^2 = 0$) (black lines), $e = 0.54$ ($\bar{\Omega}^2 = 0.54$) (blue lines), and $e = 0.69$ ($\bar{\Omega}^2 = 0.60$) (magenta lines).

In Figure 6, we compare spectra calculated for various values of the eccentricity (various dimensionless angular velocities) and for two values of the inclination angles $i = 0°$, star seen at the equator and $i = 80°$ seen almost at the pole. All these spectra are integrated over the whole surface. Spectra denoted by label 0.00 (black lines) are calculated for assumption of the spherical, nonrotating star. This figure clearly shows that spectra of spherical stars differ significantly from spectra of the distorted stars. In case of stars seen in the equatorial plane, spectra of the rotating star are harder, and their maximum is shifted toward higher energies. In contrast, the spectrum of the spherical star seen at the pole is harder than the spectra of the distorted stars.

In Figures 7 and 8, it is clearly seen that the strength of lines depends on the dimensionless angular velocity as well as on the inclination angles. For example, for $\bar{\Omega}^2 = 0.60$ ($e = 0.69$) and $i = 80°$, lines are very prominent, whereas they almost disappear for inclination angle $i = 0°$. The same effect is for our second distorted star (with $\bar{\Omega}^2 = 0.30$).

Table 2 contains values of the equivalent width and FWHM for selected Fe XXVI and Fe XXV lines. We calculated these parameters for three values of the dimensionless angular velocities $\bar{\Omega}^2 = 0.0$ (nonrotating spherical star), $\bar{\Omega}^2 = 0.30$, and 0.60 and few selected values of the inclination. We clearly show that the equivalent width and FWHM depend on the angular velocity of the rotation as well as on the inclination angle of the star.

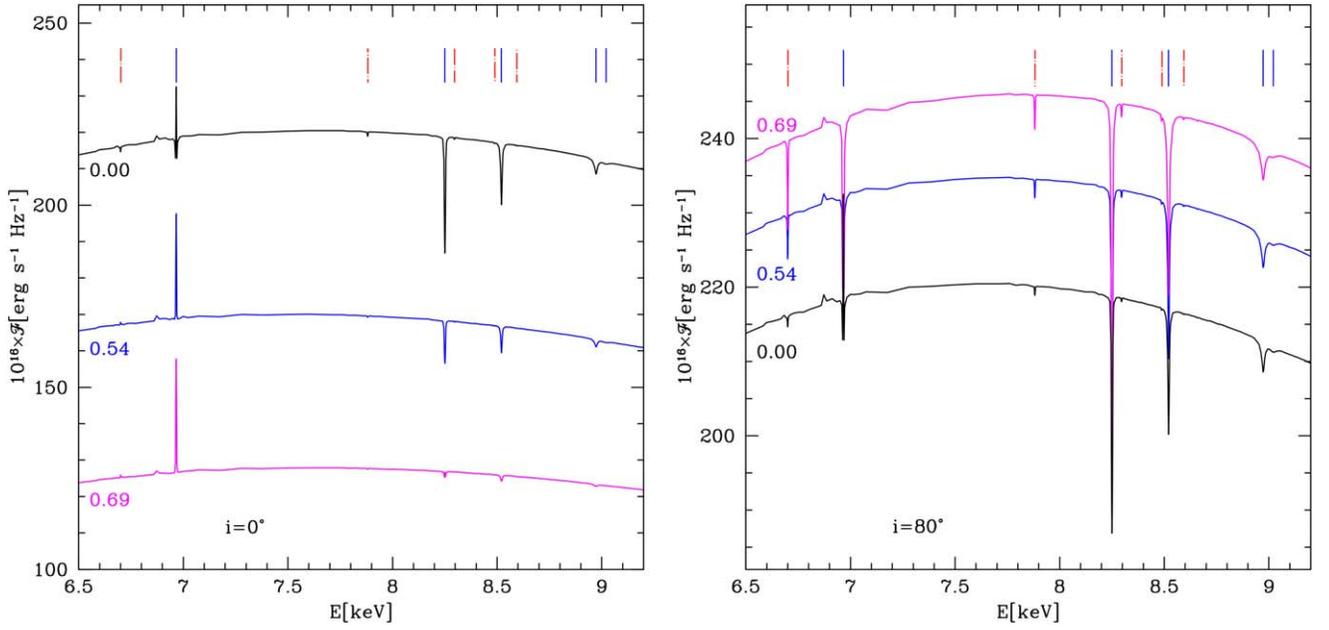

**Figure 7.** Theoretical spectra of the rotating neutron star are the same as in Figure 6. The area where spectral lines are present was magnified. By red dotted–dashed lines we mark the central energies of the fundamental series of helium-like iron, whereas blue solid lines mark spectral lines from the fundamental series of hydrogen-like iron.

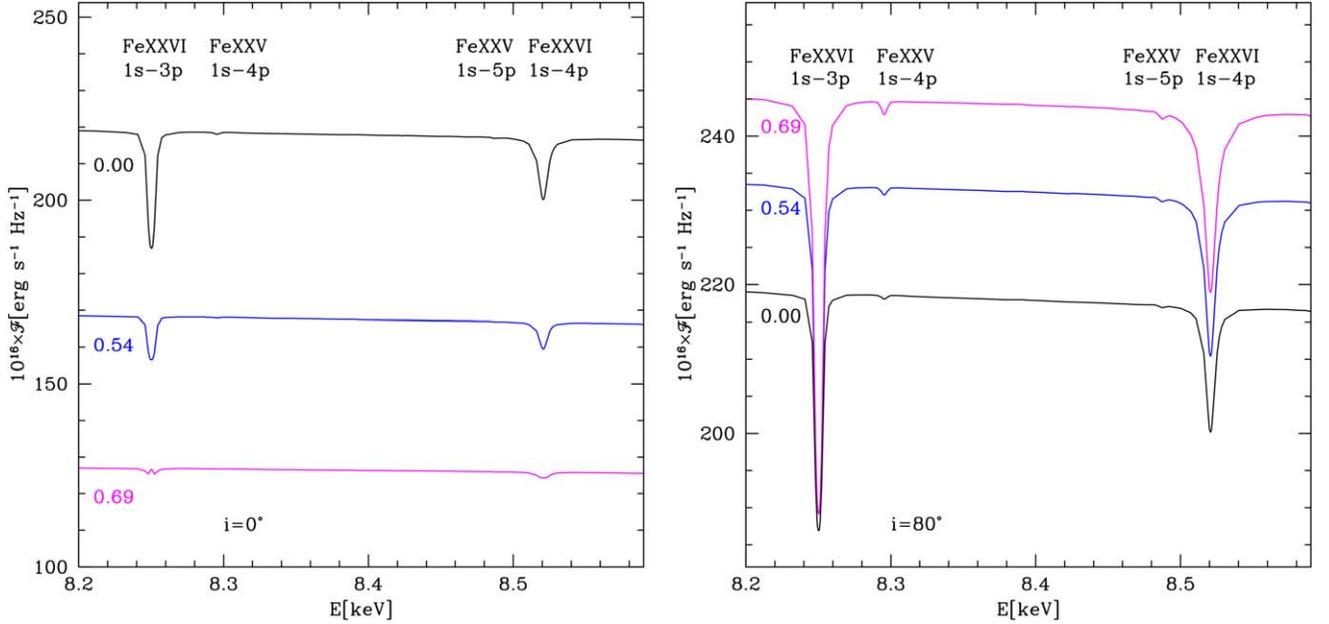

**Figure 8.** Theoretical spectra of the rotating neutron star for two assumed inclination angles $i = 0°$ (left panel) and $80°$ (right panel) and for various values of the dimensionless angular velocities $\bar{\Omega}^2 = 0.00$, 0.54, and 0.69.

## 4. Discussion and Conclusions

In this paper we presented spectra of the fast-rotating neutron stars with atmospheres containing iron. Therefore, spectral features appear in the emergent spectrum. We have shown the influence of the dimensionless angular velocity $\bar{\Omega}^2$ and the inclination angle $i$ on the shape of the spectrum and on the width of the spectral lines. For some combinations of the parameters ($\bar{\Omega}^2$ and $i$), iron lines almost disappear.

We calculated a small grid of theoretical spectra integrated over the surface of the rotationally distorted neutron star seen at various inclination angles. We assumed the following parameters: two values of the dimensionless angular velocity $\bar{\Omega}^2 = 0.30$ and 0.60, the effective temperature of the spherical star $T_{\rm eff} = 2.20 \times 10^7$ K, logarithm of surface gravity of the spherical star $\log(g) = 14.40$ (cgs), and various values of the inclination angles from $i = 0°$ to $i = 90°$ with the step $\Delta i = 10°$.

Table 2
Values of the Equivalent Width and FWHM of Fe XXVI and Fe XXV for Three Values of the Dimensionless Angular Velocity $\bar{\Omega}^2 = 0.0$, 0.30, and 0.60 and for Various Values of the Inclination

| | Central Energies of Iron Lines | | | | | | | | | |
|---|---|---|---|---|---|---|---|---|---|---|
| | Fe XXVI | | | | | | Fe XXV | | | |
| | 8250.227 | | 8520.679 | | 8972.674 | | 6700.049 | | 7881.033 | |
| $i$ | EW (eV) | FWHM (eV) | EW (eV) | FWHM (eV) | EW (eV) | FWHM (eV) | EW (eV) | FWHM (eV) | EW (eV) | FWHM (eV) |
| | $\bar{\Omega}^2 = 0.00$ | | | | | | | | | |
| ⋯ | 1.113 | 6.565 | 0.645 | 7.316 | 0.325 | 15.287 | 0.022 | 3.789 | 0.043 | 3.992 |
| | $\bar{\Omega}^2 = 0.30$ | | | | | | | | | |
| 20° | 0.675 | 6.663 | 0.426 | 7.241 | 0.237 | 14.824 | ⋯ | ⋯ | 0.030 | 4.370 |
| 40° | 1.027 | 6.752 | 0.589 | 7.575 | 0.289 | 15.971 | 0.055 | 4.002 | 0.043 | 4.194 |
| 60° | 1.366 | 6.815 | 0.734 | 7.790 | 0.332 | 16.751 | 0.097 | 3.887 | 0.056 | 4.147 |
| 80° | 1.558 | 6.845 | 0.812 | 7.884 | 0.354 | 17.081 | 0.122 | 3.870 | 0.063 | 4.132 |
| | $\bar{\Omega}^2 = 0.60$ | | | | | | | | | |
| 20° | 0.476 | 7.876 | 0.316 | 8.669 | 0.166 | 20.158 | ⋯ | ⋯ | 0.025 | 4.290 |
| 40° | 1.221 | 7.249 | 0.636 | 8.418 | 0.270 | 19.109 | 0.109 | 4.060 | 0.059 | 4.2060 |
| 60° | 1.798 | 7.208 | 0.851 | 8.449 | 0.331 | 18.797 | 0.196 | 4.069 | 0.087 | 4.2251 |
| 80° | 2.073 | 7.247 | 0.942 | 8.540 | 0.354 | 19.053 | 0.243 | 4.076 | 0.101 | 4.2408 |

We assumed that the atmosphere contained hydrogen, helium, and iron with $M_H = 0.299$, $M_{He} = 0.700$, and $M_{Fe} = 0.001$.

Fast-rotating neutron stars should be distorted; therefore, according to the von Zeipel theorem, the surface gravity and the surface temperature are not uniform over the surface of the star. We calculate local surface gravity by using a formula taken from AlGendy & Morsink (2014). The shape of the rotating neutron star depends on the properties of the matter building up the star and on the dimensionless angular velocity $\bar{\Omega}^2$. The equation of state (EOS) is in fact unknown; therefore, we used an approximate formula independent of EOS that connects the shape of the rotating star (the eccentricity $e$) with the dimensionless angular velocity $\bar{\Omega}^2$ (Silva et al. 2021). The surface effective temperature was determined from the von Zeipel law with GDE $\beta = 1/4$.

As was mentioned above, we assumed fast rotation of the neutron star; therefore, we included the quadrupole moment of the mass distribution in the star interior by using an appropriated equation on the local surface gravity (Equation (4)). We divided the star into 648 patches, and we very carefully calculated the specific intensity $I_\nu(T_{\rm eff}(\theta), g_{\rm eff}(\theta))$ at each patch. Next, these specific intensities were integrated over the whole surface of the neutron star. In this manner, we obtained the spectrum of the rotating neutron star as could be seen by the distant observer (Różańska et al. 2017, 2018). In this paper, we did not include relativistic effects of the special relativity (relativistic Doppler broadening), which changes the shape of the spectrum. Influence of this effect on the shape of the spectrum was discussed by Baubock et al. (2015).

Baubock et al. (2013) calculated profiles of the spectral lines emitted by a fast-rotating neutron star. The authors discussed the influence of the velocity of the rotation on the spectral line profiles and showed that in a fast-rotation regime, the Schwarzschild metric cannot be used. Fast-rotating neutron stars are oblate, and effects related with the quadrupole moment require other than Schwarzschild spacetime metrics; therefore, the Hartle–Thorne approximation was used (Hartle & Thorne 1968). The main conclusion of this paper is that the oblateness and the quadrupole moment need to be taken into account. In our paper, we did not include the general relativity effects, but calculation of the effective gravity includes the quadrupole moment. Neutron stars in X-ray bursters rotate with moderately velocities and are oblate; therefore, as a first approximation, we plan to add to our model line broadening due to the Doppler effect.

In this paper we show that the gravitational darkening effect strongly affects the appearance of the spectral lines. For some sets of parameters ($\bar{\Omega}^2$ and $i$), iron lines are very weak or even disappear. For example, spectra of the neutron star seen in the equatorial plane that rotates with $\bar{\Omega}^2 = 0.60$ did not show lines from helium-like iron, whereas the spectrum of the same star seen in the polar plane shows these lines (see Figure 7). This fact clearly indicates that gravitational darkening effect should be included in realistic calculations of the rotationally distorted neutron star.

## Acknowledgments

We would like to thank the anonymous referee for very helpful comments. This work was supported by grant No. 2021/41/B/ST9/04110 from the Polish National Science Center.

## ORCID iDs

Agnieszka Majczyna  https://orcid.org/0000-0003-0864-8779
Jerzy Madej  https://orcid.org/0000-0001-8417-1509
Agata Różańska  https://orcid.org/0000-0002-5275-4096
Mirosław Należyty  https://orcid.org/0000-0003-0478-5426